\documentclass{iaus}
\usepackage{graphicx,epsfig}
\usepackage{natbib}

\def\plottwo#1#2{\centering \leavevmode
\epsfxsize=.45\columnwidth \epsfbox{#1} \hfil
\epsfxsize=.45\columnwidth \epsfbox{#2}}
\def\mycite#1{\citeauthor{#1} \citeyear{#1}}

\def\aap{\rm {A\&A}}		
\def\aaps{\rm {A\&AS}}
\def\aj{\rm {AJ}}			
\def\apjl{\rm {ApJ}}		
\def\apj{\rm {ApJ}}			
\def\nat{\rm {Nature}}		

\title[The SuperMACHO Microlensing Survey]{The SuperMACHO Microlensing Survey}

\author[A. C. Becker \etal]%
{ Andrew C. Becker$^{1}$,
  A. Rest$^{2}$,
  C. Stubbs$^{3}$,
  G. A. Miknaitis$^{1}$,
  A. Miceli$^{1}$,
  R. Covarrubias$^{1}$,
  S. L. Hawley$^{1}$,
  C. Aguilera$^{2}$,
  R. C. Smith$^{2}$,
  N. B. Suntzeff$^{2}$,
  K. Olsen$^{2}$,
  J. L. Prieto$^{2}$,
  R. Hiriart$^{2}$,
  A. Garg$^{3}$,
  D. L. Welch$^{4}$
  K. H. Cook$^{5}$,
  S. Nikolaev$^{5}$,
  A. Clocchiatti$^{6}$,
  D. Minniti$^{6}$,
  S. C. Keller$^{7}$,
  \and B. P. Schmidt$^{7}$}

\affiliation{$^1$U. Washington; 
$^2$CTIO;
$^3$Harvard;
$^4$McMaster U.;
$^5$LLNL;
$^6$P. Universidad Cat\'{o}lica;
$^7$ANU.
}

\pubyear{2004}
\volume{225}
\pagerange{1--8}
\date{?? and in revised form ??}
\jname{Impact of Gravitational Lensing on Cosmology}
\editors{Mellier, Y. \& Meylan,G. eds.}
\begin{document}

\maketitle

\begin{abstract}
We present the first results from our next-generation microlensing
survey, the SuperMACHO project.  We are using the CTIO 4m Blanco
telescope and the MOSAIC imager to carry out a search for microlensing
toward the Large Magellanic Cloud (LMC).  We plan to ascertain the
nature of the population responsible for the excess microlensing rate
seen by the MACHO project.  Our observing strategy is optimized to
measure the differential microlensing rate across the face of the LMC.
We find this derivative to be relatively insensitive to the details of
the LMC's internal structure but a strong discriminant between
Galactic halo and LMC self lensing.  In December 2003 we completed our
third year of survey operations.  2003 also marked the first year of
real-time microlensing alerts and photometric and spectroscopic
followup.  We have extracted several dozen microlensing candidates,
and we present some preliminary light curves and related information.
Similar to the MACHO project, we find SNe behind the LMC to be a
significant contaminant - this background has not been completely
removed from our current single-color candidate sample.  Our follow-up
strategy is optimized to discriminate between SNe and true
microlensing.
\end{abstract}

\firstsection 
\section{Introduction}\label{sec:intro}

\cite{Paczynski86} first suggested searching for Galactic dark matter
in the form of MACHOs (MAssive Compact Halo Objects) by searching for
gravitational microlensing of stars in the Magellanic Clouds.  Several
groups followed this suggestion and established microlensing searches
that have led to a wealth of information on stellar variability and
constraints on Galactic structure.  In particular, the MACHO group
reported 13-17 microlensing events toward the LMC \citep{Alcock00a}
with event timescales (Einstein diameter crossing time $\hat{t}$)
ranging from 34 to 230 days.  This study led to an estimated
microlensing optical depth toward the LMC of $\tau =
1.2^{+0.4}_{-0.3}\times 10^{-7}$.  If we assume that MACHOs are
responsible for this optical depth, then a typical Galactic halo model
allows for a MACHO halo fraction of 20\% (95\% confidence interval of
8\%-50\%) with MACHO masses ranging between 0.15 and 0.9 $M_{\odot}$.
The EROS group found 3 LMC events (\mycite{Aubourg93},
\mycite{Renault97}, \mycite{Lasserre00}), in agreement with the
findings of the MACHO group\footnote{See however the paper by
P. Jetzer in these proceedings.  He reports long--term EROS monitoring
has detected secondary microlensing--like peaks in one previous EROS
and one previous MACHO event, suggesting stellar variability might
play a role in the observed event rate.  Jetzer also reports the most
recent EROS optical depth estimates are well below that reported by
the MACHO collaboration}.
Notably, none of the surveys toward the LMC have detected events with
timescales $1 {\rm hr} \le \hat{t} \le 10 {\rm days}$.  This complete
lack of short timescale events puts a strong upper limit on the
abundance of low-mass dark matter objects: objects with masses
$10^{-7} M_{\odot}<~M<~10^{-3} M_{\odot}$ make up less than 25\% of
the halo dark matter.  Further, less than 10\% of a standard spherical
halo is made of MACHOs in the $3.5 \times 10^{-7} M_{\odot}<~M<~4.5
\times 10^{-5} M_{\odot}$ mass range \citep{Alcock98}.  Surveys
towards M31 are also yielding a clear microlensing signal, although
the overall number of events is low and it is still difficult to
distinguish between M31 self and halo lensing (POINT-AGAPE,
\mycite{Paulin-AGAPEph04}; VATT/Columbia, \mycite{Uglesich04ph}).

Despite these constraints on the MACHO halo fraction, the reported
microlensing event rate toward the LMC significantly exceeds that
expected from known visible components of our Galaxy.  This rate
depends on the spatial, mass, and velocity distribution of the lenses.
The primary observable quantity in any given microlensing event, its
duration, depends upon a combination of all three of these parameters.
Any conclusion about the spatial location of the lens population
therefore depends upon the assumptions made about its mass and
velocity.  In cases where the lightcurve exhibits a departure from the
point-source, point-lens, inertial motion approximation (due to e.g. a
binary lens, a binary source or parallax), this degeneracy can be
lifted.

In light of the inherent difficulty in locating the LMC lenses down
the line of sight, we are left with a variety of possible explanations
for the excess LMC event rate which includes: (1) lensing by a
population of MACHOs in the Galactic Halo, (2) lensing by a previously
undetected thick disk component of our Galaxy, (3) disk-bar or bar-bar
self-lensing of the LMC, or (4) lensing by an intervening dwarf galaxy
or tidal tail.  Due to the limited number of events observed to date
it is not yet clear which scenario or combination of scenarios
explains the observed lensing.

Each of these dominant lens populations produces a signature in the
microlensing optical depth as a function of position across the face
of the LMC.  Using recent papers by \cite{vdMarel01a},
\cite{vdMarel01b}, and \cite{vdMarel02} on the structure and
kinematics of the LMC, \cite{Mancini04ph} derive and compare these
spatial signatures for a variety of lens populations.  In particular,
Galactic halo microlensing should produce a slight optical depth
gradient across the face of the LMC, varying by $\sim 6\%$.  LMC
self--lensing tends to produce a steep gradient in optical depth
around a central peak.  However, the reported MACHO and EROS events
only coarsely sample these optical depth contours.  To do so finely,
and in turn constrain the nature of the lensing populations(s),
requires an order of magnitude increase in the number of detected
events.

\section{The SuperMACHO Survey}\label{sec:sm}

The SuperMACHO
Project\footnote{http://www.ctio.noao.edu/$\sim$supermacho} is an
ongoing five-year microlensing survey of the LMC that is being carried
out with the specific goal of determining the location of the objects
that produce the observed microlensing events \citep{Stubbs99}.  This
will be achieved by adding more events to the overall sample, allowing
comparisons/distinctions to be drawn between the measured optical
depth variation and Galactic--LMC lensing models.  We have been
allocated 150 half-nights, distributed over 5 years, on the Cerro
Tololo Interamerican Observatory (CTIO) Blanco 4m telescope through
the NOAO Survey Program.  The survey started in 2001 and will run
through 2005
\footnote{We note that we have waived proprietary data access rights,
and that the SuperMACHO survey images are accessible through the NOAO
Science Archive on the NOAO web site at
ftp://archive.tuc.noao.edu/pub/}.

Observations are carried out every other night in dark time during the
months of October, November, and December, when the LMC is most
accessible from CTIO.  We use the $8K\times8K$ MOSAIC II CCD imager
with a FOV of 0.33 square degree.  The 8 SITe $2K\times4K$ CCDs are
read out in dual-amplifier mode (i.e.  different halves of each CCD
are read out in parallel through separate amplifiers) to increase our
observing efficiency.  In order to maximize the throughput we use a
custom-made broadband filter (VR filter) from 500nm to 750nm.  The
atmospheric dispersion corrector on the Mosaic imager allows for the
use of this broad band without a commensurate PSF degradation.  We
monitor 68 LMC fields, using difference image analysis to search for
variability.  Our combination of exposure times and telescope aperture
means that we are sensitive to changes in brightness equivalent to the
brightness of a $\sim 23^{rd}$ magnitude star.  \cite{Rest04} estimate
that we are sensitive to this level of variability in 50--100 million
stars.

Our data are reduced immediately after acquisition using an automated
data reduction pipeline.  Images are cross--talk corrected and WCS
calibrated as a Multi--Extension Fits file, and afterwards separated
into 16 separate images, one for each amplifier, for further parallel
processing.  The images are calibrated, their PSFs are modeled using a
modified version of the {\tt DOPHOT} program, and are then passed to
an image registration module (currently based on the
SWarp\footnote{http://terapix.iap.fr/soft/swarp} program) that
astrometrically aligns each with a reference template image.  The
input and template images are then differenced using a modified
version of the \cite{Alard00} algorithm \citep{Becker04a}.  {\tt
DOPHOT} is used to detect new objects using the PSF derived from the
input images.  For any new objects detected, forced--position
photometry is applied to the entire time--series of images containing
the position of the variability.  Cuts are applied to these time
series to look for true variability, and candidate events passing
these cuts are posted to an internal web page for visual review.
Events passing this stage are posted to a public web page.

\subsection{LMC Fields}

\begin{figure}
  \plottwo{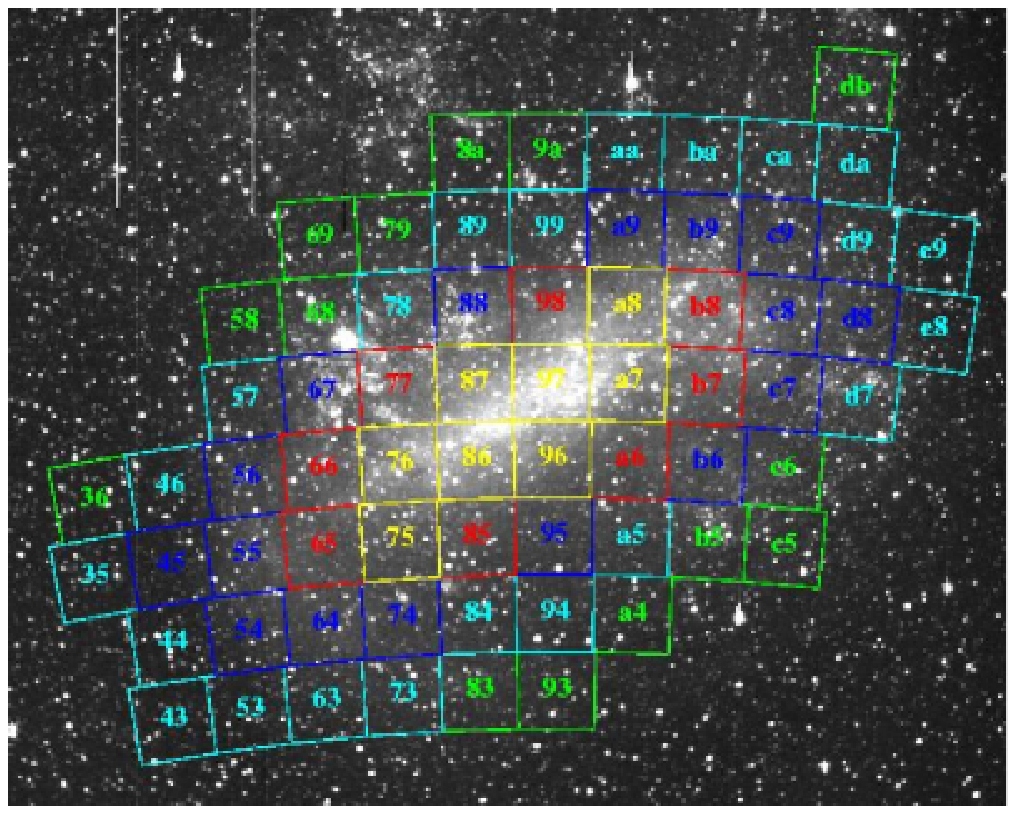}{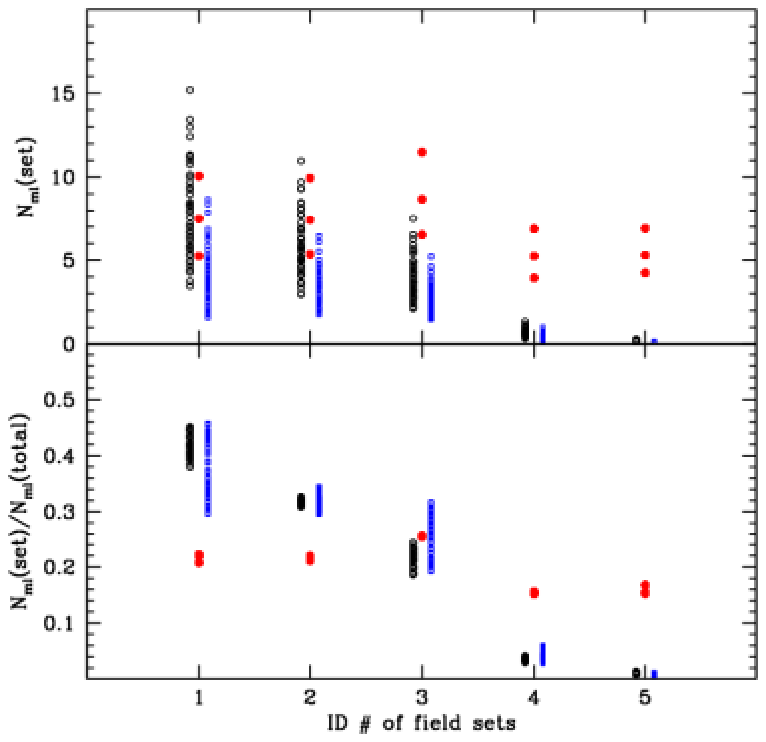}
  \caption{{\it Left} : Distribution of SuperMACHO fields across the
    face of the LMC.  For the purpose of analysis, these fields are
    divided into sets comprising roughly equal numbers of source
    stars.  Field sets 1--5 are shown in {\it yellow}, {\it red}, {\it
    dark blue}, {\it light blue}, and {\it green}, respectively, and
    are arranged in concentric rings around a fiducial center.  {\it
    Right} : Cumulative and differential distribution of microlensing
    events across the field sets, for 3 different microlensing models.
    For each field set, the {\it black} points represent Zhao \& Evans
    LMC self--lensing models, the {\it red} are for Galactic halo
    lensing assuming the MACHO results, and the {\it blue} are for
    Nikolaev LMC self--lensing.  The distributions are ordered as
    above and horizontally offset for clarity.  Uncertainly in the
    expected number of events comes from varying model parameters.
    The bottom panel shows that the differential event rate is less
    sensitive to these parameters.  In particular, field sets 4 and 5
    are highly sensitive to Galactic halo MACHOs.}\label{fig:fields}
\end{figure}

Figure~\ref{fig:fields} shows our distribution of fields on the LMC.
For the purpose of our analysis, we have arranged these into 5 sets of
fields, each of which contain roughly the same number of LMC source
stars.  
Figure~\ref{fig:fields} also shows the expected cumulative and
differential event distribution amongst these fields sets.  For each
set, we show the expected number of events for 3 lensing model
populations.  The black points show the number and fraction of events
expected from the LMC self--lensing model of \cite{Zhao00a}, the red
points are for Galactic halo lensing adopting the \cite{Alcock00a}
results, and the blue for the LMC self--lensing model of Nikolaev
(2003, private communication).  In addition, model parameters such as
the LMC luminosity function (LF), disk--bar separation, and bar mass
fraction are varied to add a dispersion in the expected number of
events that reflects our uncertainty in the physical system.  The
lower panel shows the fraction of all microlensing events detected in
each set.  The smaller dispersion indicates this metric is much less
sensitive to the model parameters while still retaining sensitivity
between self and Galactic halo lensing models.  If the MACHO fraction
is significantly lower than the \cite{Alcock00a} results (P. Jetzer,
these proceedings; A. Milsztajn, private communication), the
cumulative number of events expected will change, but the differential
rate should not.  In addition, field sets 4 and 5 are almost uniquely
sensitive to the Galactic halo MACHO fraction, avoiding complications
which might result from an admixture of lensing populations.

The lowest Galactic halo lensing expectation value for the field set 5
differential event rate is 0.15.
As a particular test to exclude LMC self--lensing models, and assuming
the total and fractional number of events expected from the above halo
lensing scenario, we next note the upper limit on the field set 5
differential measurement for all Zhao self--lensing models is 0.014.
We can exclude these models with 99\% and 99.5\% confidence if we
observe a differential rate larger than 0.11 and 0.13, respectively.
Similarly for the Nikolaev models, the maximum differential rate is
0.011, thus they can be excluded at 99.5\% confidence if the observed
differential event rate is bigger than 0.12.  Galactic halo lensing is
highly favored if the measured field set 5 event fraction is $\gtrsim
0.1$.

\subsection{Event Candidates : 2003}

\begin{figure}
  \plottwo{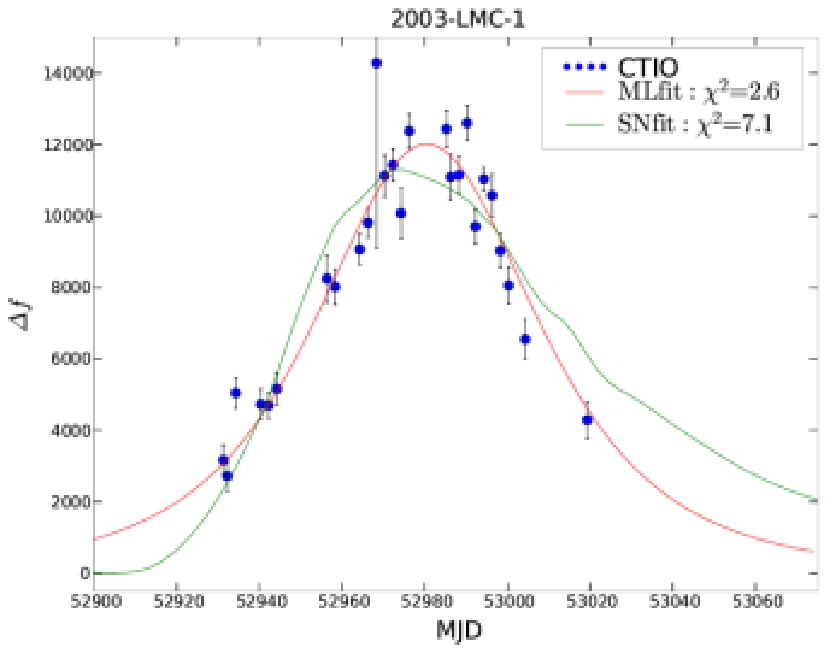}{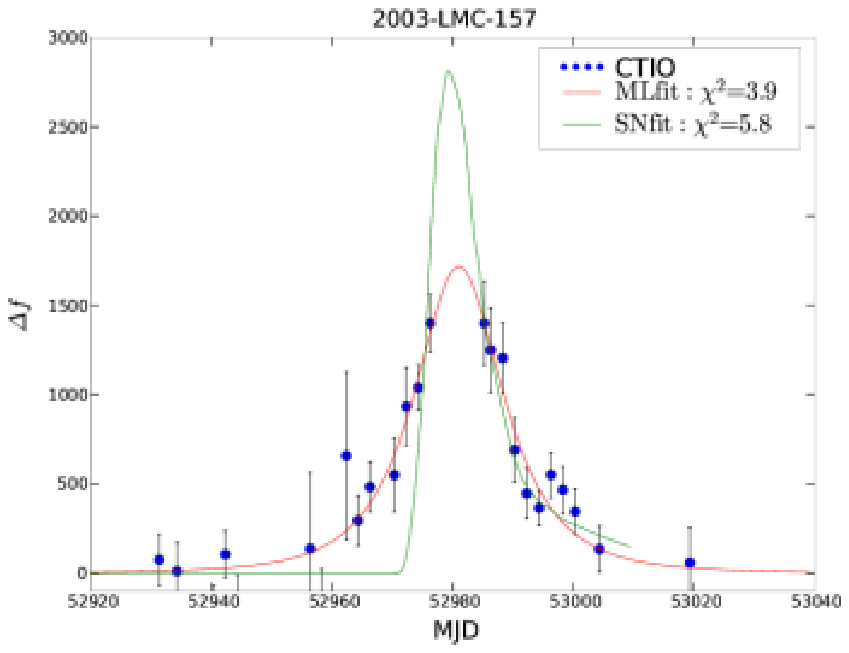}
  \caption{Two microlensing candidates from 2003.  2003--LMC--1 was a
    low magnification event ($u_{min} = 1.5$) on a likely clump--giant
    star, and yielded a peak magnitude of $VR = 18.9$.  2003--LMC--157
    is a more typical event.  The lensed source star is unknown, as
    there is no obvious counterpart in our template image.  We can
    limit the impact parameter to be $u_{min} < 0.8$.  The observed
    peak change in flux yields $\Delta VR = 23.1$.}
  \label{fig:events}
\end{figure}

\begin{figure}
  \plottwo{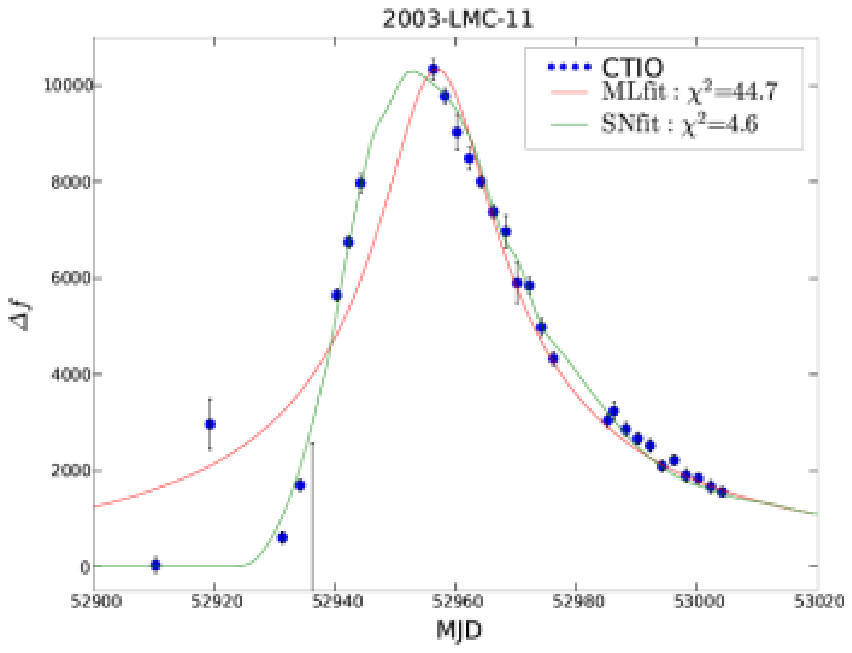}{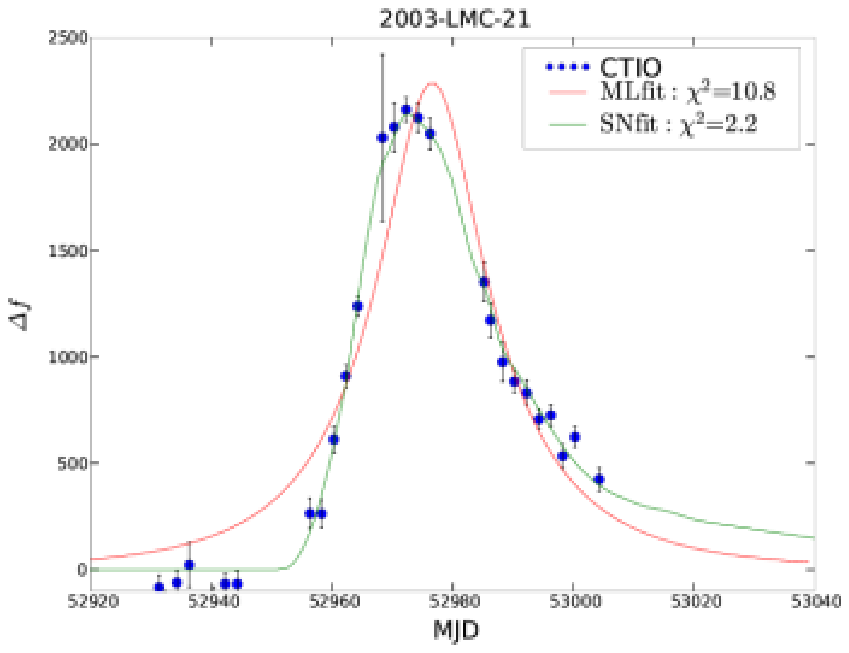}
  \caption{Two supernova candidates from 2003.  We obtained
    spectroscopy of event 2003--LMC--11, indicating a Type Ia SN at a
    redshift of $z = 0.25$.  2003--LMC--21 is determined to be a
    likely SN due to the good fit to SNe lightcurve templates compared
    to the best microlensing fit ($\Delta \chi^2 = 8.6$) and the clear
    lightcurve asymmetry.}
  \label{fig:sn}
\end{figure}

We detected approximately 10 high quality microlensing events in 2003,
two of which are shown in Figure~\ref{fig:events}.  We also found
nearly 70 supernovae (SNe) in galaxies behind the LMC, two of which
are shown in Figure~\ref{fig:sn}.  These are frequently
distinguishable by the presence of their host background galaxy in our
images.  At high signal--to--noise (S/N), the lightcurves are also
clearly asymmetric, in the particular sense that the rising slope is
steeper than the falling slope.  At low S/N, however, it is more
difficult to distinguish the two populations of events.  We note that
in 2003 we also obtained spectroscopy of approximately 25 microlensing
candidates, which is used to both confirm and reject particular events
as microlensing.  Spectroscopy during the event is arguably the most
discriminating observation, and we regard this campaign as an integral
part of the survey.

\subsection{SuperMACHO 2004--2005}

For the upcoming observing seasons, we will be modifying our observing
strategy and allocating photometric and spectroscopic resources to
define concretely the characteristics of microlensing and supernovae
in our data.  Our alert and followup routine will take the following
form :

\begin{itemize}
\item Alert based upon flat prior baseline, rising flux, and other
  characteristics of the time--series.  Based upon our experiences in
  2003, we are confident we can detect variability well before peak.
\item 1st SN discriminant : Presence/absence of host galaxy in
  template image.
\item 2nd SN discriminant : Multi-color observations from the 4--m
  before peak.  $VR-I$ vs $B-VR$ colors will yield a distance from the
  stellar locus as well as an early reference point for color
  evolution studies.
\item 3rd SN discriminant : Snapshot high--resolution imaging from
  Gemini South or Magellan.  This imaging will also serve as a finding
  chart and/or resource for nod--and--shuffle planning for future
  spectroscopy, if needed.  This will also yield an additional check
  for background galaxies, as well as another epoch of color
  information in which the sources are less blended.
\item Objects passing the above cuts are considered good candidates
  for microlensing, and will be submitted to Gemini South or Magellan
  for spectroscopic followup near peak.
\item For those events where we are unable to obtain spectroscopy, we
  will instead obtain additional multi--color images during the event,
  especially near peak, to constrain color evolution.  The
  color--evolution of SNe is 2 magnitudes in $VR-I$ before the peak,
  and nearly 2 magnitudes in $B-VR$ afterwards.  For microlensing, the
  color of the variability should remain in place on the stellar
  locus.  In addition, SNe $VR-I$ colors at peak are significantly
  removed from the stellar locus, and serve as a powerful
  single--epoch discriminant.  This photometric followup will occur
  using the 4--m and other global resources.
\end{itemize}


\section{Conclusions}\label{sec:concl}

The SuperMACHO survey intends to constrain the location of the objects
contributing to the excess microlensing signal towards the LMC.  By
increasing the overall number of events by a factor of several over
the current event count, we will be able to more faithfully sample the
variation of the microlensing optical depth across the face of the
LMC, and thus constrain the nature of the lenses yielding this signal.
As of 2003, the survey is detecting and following up events in
real--time.  We find that background supernovae are our most prominent
source of contamination.  Our survey and follow--up strategy are
designed to obtain multi--color and spectroscopic information of
on--going events to discriminate true microlensing from the sources of
background.  Spectroscopic information is necessary to absolutely
confirm the normal stellar nature of candidate microlensing sources,
and to rule out intrinsic variability as the source of a detected
event.

\begin{acknowledgments}

The SuperMACHO survey is being undertaken under the auspices of the
NOAO Survey Program. We are very grateful for the support provided to
the Survey program from the NOAO and the National Science Foundation
We are particularly indebted to the scientists and staff at the Cerro
Tololo Interamerican Observatory for their assistance in helping us
carry out the SuperMACHO survey. 
The support of the McDonnell Foundation, through a Centennial
Fellowship awarded to C. Stubbs, has been essential to the SuperMACHO
survey. 
KHC's and SN's work was performed under the auspices of the U.S.
Department of Energy, National Nuclear Security Administration by the
University of California, Lawrence Livermore National Laboratory under
contract No. W-7405-Eng-48.
DLW acknowledges financial support in the form of a Discovery Grant
from the Natural Sciences and Engineering Research Council of Canada
(NSERC).
\end{acknowledgments}


\end{document}